\documentclass[aps,amsmath,amssymb,amsfonts,tightenlines,prd,10pt,onecolumn,preprint,superscriptaddress,nofootinbib]{revtex4-1}

\usepackage{bm, slashed}
\usepackage{graphicx}
\usepackage{epic,eepic}
\usepackage{axodraw4j}
\usepackage{enumitem}

\newcounter{incre}
\newcommand{\benum}{\begin{enumerate}[label=\roman{enumi}),ref=\roman{enumi}] \setcounter{enumi}{\value{incre}}}
\newcommand{\eenum}{\setcounter{incre}{\value{enumi}}\end{enumerate}}

\newcommand{\ArcTan}{\mbox{Tan}^{-1}}

\begin{document}

\title{Flavor Oscillation from the Two-Point Function}
\author{Mario Martone}
\email{mcm293@cornell.edu}
\affiliation{Laboratory for Elementary-Particle Physics, Cornell University, Ithaca, N.Y.}
\affiliation{Dipartimento di Scienze Fisiche, University of Napoli and INFN,
Via Cinthia I-80126, Napoli}
\affiliation{Department of Physics,
Syracuse University, Syracuse, N.Y.}
\author{Dean J. Robinson}
\email{djr233@cornell.edu}
\affiliation{Laboratory for Elementary-Particle Physics, Cornell University, Ithaca, N.Y.}
\date{\today}

\begin{abstract}
We present a formalism for the flavor oscillation of unstable particles that relies only upon the structure of the time Fourier-transformed two-point Green's function. 
We derive exact oscillation probability and integrated oscillation probability formulae, and verify that our results reproduce the known results for both neutrino and neutral meson oscillation in the expected regimes of parameter space.  The generality of our approach permits us to investigate flavor oscillation in exotic parameter regimes, and present the corresponding oscillation formulae.
\end{abstract}

\maketitle

\section{Introduction}
The phenomenon of flavor oscillation plays an important role in the physics of neutral meson and neutrino systems. In particular, flavor oscillation provides the only means to measure the extremely small mass and decay rate splittings among the neutral mesons, and also provides convincing evidence for the existence of non-zero neutrino masses. The theoretical descriptions of flavor oscillation fall into several categories, including the basic plane wave Pontecorvo formalism \cite{Pontecorvo:1968, Bilenky:1978lm}, intermediate \cite{Kayser:1981qm,Giunti:1991wd,Giunti:1992rw,Lipkin:1995tn,Kiers:1995zj,Grossman:1996eh,Lipkin:1999nb,Giunti:2000kw,Giunti:2002xg,Beuthe:2002ej,Bilenky:2011pk} and external \cite{Rich:1993qm, Giunti:1993se, Grimus:1996av, Giunti:1997sk, Campagne:1997no, Grimus:1998ft, Ioannisian:1999no, Stockinger:2000if} wavepacket approaches and quantum field theoretic results \cite{Okun:1982sr, Alfinito:1995sn, Blasone:1999ef, Binger:1999nj, Blasone:2001du, Blasone:2002jv, Li:2006qt, Cohen:2009dn, Keister:2010rq, Dvornikov:2010dc}. Some detailed reviews of these approaches, their underlying assumptions, results and difficulties can be found in Refs. \cite{PDG:2010,Bilenky:1978lm, Beuthe:2001rc, Akhmedov:2010ms} (and references therein). To be very brief: In the first, one assumes the flavor states are unitary combinations of plane-wave mass eigenstates that follow spacetime worldlines, and one must carefully define the proper times of the mass eigenstates in order to obtain the well-known Pontecorvo oscillation formula. The intermediate wavepacket approach treats the oscillating degrees of freedom as a linear combination of one-particle states, while the external wavepacket approach treats the oscillating particles as quantum fields, whose propagator is convolved with wavepackets at the source and detector.

A large amount of Literature has been devoted to deriving, studying and comparing oscillation formulae within these different approaches. Particular care has been taken to include important effects such as measurement uncertainties, coherence effects, the finite size of the detector and source, all of which together lead to somewhat complicated formulae. Our goal in this paper is less ambitious: Using a quantum field theoretic approach, we present a simple formalism of oscillation based entirely on the properties of the \emph{spatially} Fourier transformed propagator. We call this the spatial two-point function. The resulting oscillation formulae are particularly elegant, and precisely reproduce both the Pontecorvo neutrino result and (CP violating) neutral meson mixing results (see e.g. Ref. \cite{Branco:1999cp,Bigi:2009cp}) in appropriate parameter regimes. 

To construct this formalism, we assume that the oscillation experiment measures the exchanged energy $E$ and source-detector displacement $\bm{L}$ to infinite precision, along with flavor at both the source and detector. This is possible because $E$, $\bm{L}$ and flavor are commuting observables, so that an amplitude which depends exclusively on these quantities is well-defined. The key idea is that the spatial two-point function $\Delta(E,\bm{L})$ in the flavor basis is a well-defined amplitude which encodes flavor oscillation over a displacement $\bm{L}$ at energy $E$. We therefore assume that the experiment amplitude is proportional to $\Delta(E,\bm{L})$ and explore the resulting oscillation formulae. The advantages of this description are: There is no ambiguity in the choice of reference frame - all computations are done in the lab frame and one never needs to introduce proper times into the formalism; the oscillation probabilities can be computed exactly; and one obtains formulae whose physical meaning can be easily discerned in various limits. Since we do neglect several real physical effects mentioned above --- in particular the physics of the source and detector are neglected --- the limits of the applicability of our theoretical description to actual oscillation experiments should be carefully examined. Nonetheless, we believe this approach provides an instructive, leading order description of the physics of oscillation in real experiments.

In terms of the previous Literature on this subject, our approach is best categorized as a special case of the above-mentioned external wavepacket formalism with stationary states \cite{Grimus:1996av}. However, to our knowledge, the oscillation physics contained just in the two-point function has not been thoroughly investigated and the resulting general oscillation formulae for unstable particles obtained by our approach have not been previously presented.  One exception is Ref. \cite{Okun:1982sr}, whose amplitudes for stable fermions agree with our results for the special case of stable particles.

This paper is structured as follows. In Section \ref{sec:F} we present the oscillation formalism. In Section \ref{sec:EOP} the exact spatial two-point function $\Delta(E,\bm{L})$ for unstable fields is presented, and the exact oscillation probabilities and integrated oscillation probabilities are computed. In Section \ref{sec:R} we examine our results in several different parameter regimes and recover both the neutral meson-mixing results and Pontecorvo neutrino oscillation results in appropriate limits. 

\section{Formalism}
\label{sec:F}
\subsection{Experiment Amplitude}
\label{sec:EA}
Our starting point is to consider an experiment which involves the propagation between a source and a detector of a set of fields $\{\phi^\alpha\}$, which are allowed to mix. As usual, $\alpha$ is an experimentally measurable label called the flavor, which is henceforth always denoted by a Greek index. The set $\{\phi^\alpha\}$ is called the flavor field basis.

In this paper we assume the $\alpha \to \beta$ oscillation experiment measures the exchanged energy $E$ and source-detector displacement $\bm{L}$ to infinite precision in the lab frame. The amplitude for the experiment must then have the form
\begin{equation}
	\label{eqn:FFA}
	\mathcal{M} = \mathcal{M}_{\alpha\beta}(E,\bm{L})~.
\end{equation}
This is a well-defined amplitude since $E$, $\bm{L}$ and flavor are commuting observables. Note that as a consequence of the infinitely precise $E$ and $\bm{L}$ measurement neither the time of travel nor the three-momentum between the source and detector is well-defined, because these observables do not commute with $E$ and $\bm{L}$ respectively. In other words the initial and final states of this amplitude must be energy-spatial eigenstates, rather than momentum-time eigenstates. 

The key idea of this paper rests on the observation that the time Fourier-transformed time-ordered exact two-point function --- the spatial two-point function --- defined by
\begin{equation}
	\Delta_{\alpha\beta}(E,\bm{L}) \equiv \int dt \Big\langle T\Big\{\phi^{\beta}(t,\bm{L})\phi^{\alpha\dagger}(0,\bm{0})\Big\}\Big\rangle e^{iEt}~,
\end{equation}
is the field theoretic object which encodes the oscillation of flavor $\alpha \to \beta$ over a displacement $\bm{L}$ with energy $E$. (As usual $\langle T\{\phi^{\beta}(x)\phi^{\alpha\dagger}(y)\}\rangle \equiv \Delta_{\alpha\beta}(x-y)$ is a function of $x-y$ due to translation invariance.) It is therefore natural to write
\begin{equation}
	\label{eqn:DEA}
	\mathcal{M}_{\alpha\beta}(E,\bm{L}) = \mathcal{A}^\alpha_{\textrm{S}}\mathcal{A}^\beta_{\textrm{D}}\Delta_{\alpha\beta}(E,\bm{L})~,
\end{equation}
(no sum over $\alpha$, $\beta$) where $\mathcal{A}^{\alpha}_{\textrm{S},\textrm{D}}$ encode the physics of the source and detector, which we have assumed factorizes out of the amplitude\footnote{The general criteria under which such a factorization may be possible in real oscillation experiments has been examined previously in detail (see e.g. Ref. \cite{Akhmedov:2010ms}).  Foregoing such a discussion, our intent here is that the physics of the source and detector can be neglected up to their ability to distinguish flavor.}. Assuming $\mathcal{A}^{\alpha}_{\textrm{S},\textrm{D}}$ are known, the implication of Eq. (\ref{eqn:DEA}) is that $|\Delta_{\alpha\beta}(E,\bm{L})|^2$ is a measurable quantity, from which we may proceed to construct oscillation probabilities.

So far we have not specified the spin of $\phi^\alpha$. As is well-known, if $\phi^\alpha$ are massive they must create spin-$j$ particles, with $j$ a half-integer, which have $2j+1$ spin degrees of freedom. We assume that these spin degrees of freedom decouple, so that we need only consider scalar propagators henceforth. 

\subsection{Oscillation Probability}
Having written down the amplitude for the experiment, we now define the flavor oscillation probability via
\begin{equation}
	\label{eqn:OPD}
	P_{\alpha \to \beta}(E,\bm{L}) \equiv \frac{\big|\Delta_{\alpha\beta}(E,\bm{L})\big|^2}{\underset{\gamma}{\sum}\big|\Delta_{\alpha\gamma}(E,\bm{L})\big|^2}~.
\end{equation}
Here $P_{\alpha\to\beta}$ forms a well-defined probability distribution, since $P_{\alpha\beta} \ge 0$ and $\sum_\beta P_{\alpha\beta} = 1$. 
In some experiments, measurement of  $|\Delta_{\alpha\beta}|^2$ at a precise $\bm{L}$ is replaced by a volume-averaged measurement, 
\begin{equation}
	\mathcal{A}^\textrm{I}_{\alpha\beta}(E) \equiv \int d^3\!\bm{L} \big|\Delta_{\alpha\beta}(E,\bm{L})\big|^2~. 
\end{equation}
This is equivalent to the time-averaged amplitudes measured in e.g. meson mixing experiments, in which the initial and final flavor states are determined by tagging via decay products (see e.g. Ref. \cite{Branco:1999cp,Bigi:2009cp}). We can correspondingly define an integrated oscillation probability
\begin{equation}
	\label{eqn:IOPD}
	P_{\alpha \to \beta}^{\textrm{I}}(E) \equiv \frac{\mathcal{A}^\textrm{I}_{\alpha\beta}(E)}{\underset{\gamma}{\sum}\mathcal{A}^\textrm{I}_{\alpha\gamma}(E)} =\frac{ \int d^3\!\bm{L}\big|\Delta_{\alpha\beta}(E,\bm{L})\big|^2}{\underset{\gamma}{\sum}\int d^3\!\bm{L}\big|\Delta_{\alpha\gamma}(E,\bm{L})\big|^2}~.
\end{equation}
This is also a well-defined probability distribution.

\subsection{Propagator and 1PI Basis}
So far in this paper we have formulated a description of flavor oscillation in terms of just the exact quantum amplitude $\Delta(E,\bm{L})$, which is equivalently defined as the spatial Fourier transform of the exact propagator, $\Delta(p^2)$. Explicitly, 
\begin{equation}
	\label{eqn:EDSTP}
	\Delta_{\alpha\beta}(E,\bm{L}) = \int \frac{d^3\bm{p}}{(2\pi)^3}\Delta_{\alpha\beta}(p^2)e^{i\bm{p}\cdot\bm{L}}~.
\end{equation}
Applying external field methods to the path-integral formulation of quantum field theory, it is a well-known result (see e.g. \cite{Weinberg:1996vol2,Zinn-Justin:2002}) that for a set of $N$ fields $\{\phi^\alpha\}$ the exact two-point function is the inverse of the exact two-point one-particle-irreducible (1PI) function: $\Delta_{\alpha\beta}(x-y) = -\Pi^{-1}_{\alpha\beta}(x-y)$. The Fourier transform of this result is
\begin{equation}
	\label{eqn:EP}
	\Delta_{\alpha\beta}(p^2) = \bigg[\frac{i}{p^2\bm{1} - M^2(p^2)}\bigg]_{\alpha\beta}~.
\end{equation}
Henceforth we shall call the $N\times N$ matrix of functions $M^2(p^2)$ the exact two-point 1PI function. 

In general, one cannot compute the exact propagator $\Delta(p^2)$ exactly for all $p^2$. However, the combination of Eqs (\ref{eqn:EDSTP}) and (\ref{eqn:EP}) suggests that the exact spatial two-point function is sensitive only to the pole structure of $\Delta(p^2)$. As we shall see below, with suitable assumptions this pole structure depends only on physical masses and rest frame decay rates, permitting us to construct exact oscillation probabilities in terms of just these measureable quantities, $E$, $\bm{L}$, and a mixing matrix, despite our incomplete knowledge of the exact propagator.

Now, the exact propagator (\ref{eqn:EP}) is generally not diagonal in flavor space --- there would be no oscillation if this were the case --- but the analytic structure of $\Delta(p^2)$ is greatly simplified if the exact propagator can be diagonalized. Ultimately, we want to be able to write
\begin{equation}
	\label{eqn:GDU}
	\Delta_{\alpha\beta}(p^2) = U^{\alpha j} (U^{-1})^{j\beta}\Delta_{j}(p^2)~,\qquad \Delta_j(p^2) \equiv \frac{i}{p^2 - M_j^2(p^2)}~,
\end{equation}
so that
\begin{equation}
	\label{eqn:CBD}
	\Delta_{\alpha\beta}(E,\bm{L}) = U^{\alpha j}(U^{-1})^{j\beta}\Delta_j(E,\bm{L})~,\qquad \Delta_j(E,\bm{L}) \equiv \int \frac{d^3p}{(2\pi)^3} \frac{ie^{i\bm{p}\cdot  \bm{L}}}{p^2 - M_j^2(p^2)}~.
\end{equation}
In Eq. (\ref{eqn:GDU}), $U$ is the constant and possibly unitary matrix that diagonalizes $\Delta(p^2)$ (equivalently $M^2(p^2)$), and $M^2_j(p^2)$ are the $N$ eigenvalues of $M^2(p^2)$. Below we'll see that the $M_j(p^2)$ determine the physical masses and rest frame decay rates of the particles propagating in $\Delta_{\alpha\beta}(p^2)$.

In constrast to the usual diagonalization of the classical Lagrangian mass terms, diagonalization of the exact propagator may be non-trivial. In Appendix \ref{sec:AB} we discuss the details of the diagonalization of $\Delta(p^2)$, the properties of $U$ and how this exact quantum formalism both relates to and differs from the usual classical mixing matrix formalism. For our purposes here, we assume $\Delta(p^2)$ is diagonalizable in the manner of Eq. (\ref{eqn:GDU}). Unless otherwise stated, we also assume $U$ is unitary. An immediate consequence of unitarity is that spatial two-point function can now be written as
\begin{equation}
	\Delta_{\alpha\beta}(E,\bm{L}) = U^{\alpha j}U^{\beta j *}\Delta_j(E,\bm{L})~.
\end{equation}

Let us now define the 1PI basis. This basis is a generalization of the mass basis derived in the classical formalism, that may accommodate both unstable particles and a description of CP violation for two flavors. In particular, if $U$ is constant (but not necessarily unitary), then there exists a well-defined second basis of fields $\{\phi^j\}$, henceforth denoted by a Latin index, which are defined via the linear transformations
\begin{equation}
	\label{eqn:UT}
	\phi^{\alpha\dagger} = U^{\alpha j}\phi^{j\dagger}~.
\end{equation}
Note that $\phi^\dagger$ creates a particle state, while $\phi$ creates an anti-particle state: We have chosen this definition of basis change by $U$ in order that it coincides with the usual definition in terms of one-particle quantum states. We call $\phi^j$ the 1PI basis for the following reason. If $U$ is unitary, then observe that not only $M^2(p^2)$ but also the two-point function is diagonal, i.e. $\langle T \{\phi^{i}(x)\phi^{j\dagger}(y)\} \rangle = \delta_{ij}\Delta_j(x-y)$. This implies that $M^2_j(p^2)$ is the 1PI function for $\phi^j$, whence the name. In contrast, if $U$ is not unitary, then even though $M^2(p^2)$ is still diagonalized by $U$, we have $\langle T \{\phi^{i}(x)\phi^{j\dagger}(y)\} \rangle \not= \delta_{ij}\Delta_j(x-y)$. Therefore $M_j^2(p^2)$ is no longer the 1PI function for $\phi^j$. Nonetheless, we shall always refer to the field basis defined by Eq. (\ref{eqn:UT}) to be the 1PI basis, and often call $\phi^{j\dagger}$ the 1PI states.

Tying the diagonalization of the exact propagator and the definition of 1PI basis together, we can now explain why we have taken care to consider the case of non-unitary $U$: We do so to accommodate CP-violating two-flavor neutral meson oscillations (for three or more flavors, even unitary $U$ may have a CP-violating phase), which we consider in Sec. \ref{sec:NUM}. The idea is that the Hamiltonian for such a system is diagonalized by a constant non-unitary matrix \cite{Bigi:2009cp,Branco:1999cp}, so we therefore expect $U$ to be non-unitary too. In this context the flavor field basis (1PI basis) then corresponds to the CP conjugate states (evolution eigenstates). One deduces $U$ is a particular constant $2\times 2$ non-unitary matrix, from which we can immediately derive the usual oscillation formulae.

\subsection{Exact Propagator Analytic Structure}
Let us finally examine the analytic structure of $\Delta_j(E,\bm{L})$, which is explicitly
\begin{equation}
	\label{eqn:1PIES}	
	\Delta_j(E,\bm{L}) = \int \frac{d^3p}{(2\pi)^3} \frac{ie^{i\bm{p}\cdot  \bm{L}}}{p^2 - M_j^2(p^2)}~.
\end{equation}
If $\phi^j$ is unstable, then the propagator $\Delta_j(p^2)$ will have a unique Breit-Wigner or resonance pole, which by convention is a simple pole located at
\begin{equation}
	\label{eqn:GEP}
	p^2 = m_j^2 -im_j\Gamma_j~.
\end{equation}
Here $m_j$ is the physical mass and $\Gamma_j \ge 0$ is the rest frame decay rate. The non-zero imaginary part for this pole enforces the usual Feynman pole prescription and associated time-ordering, so that we need not add the usual $i\epsilon$ convergence term in the denominator of Eq. (\ref{eqn:1PIES}), provided we assume $\Gamma_j \not= 0$. Consequently, taking the $\Gamma_j \to 0^+$ limit, which corresponds to $\phi^j$ being stable, can only be performed after all integrations and other limits are evaluated.  

By definition there are no higher order poles in $\Delta_j(p^2)$ and the residue at the pole (\ref{eqn:GEP}) is unity: Eq. (\ref{eqn:GEP}) and this latter condition are equivalent to $M^2_j(m_j^2 - im_j\Gamma_j) =  m_j^2 - im_j\Gamma_j$ and $M^{2\prime}_j(m_j^2 - im_j\Gamma_j) = 0$ respectively.

\section{Exact Oscillation Probability}
\label{sec:EOP}
\subsection{Spatial Two-Point Function}
Computation of the oscillation probabilities (\ref{eqn:OPD}) and (\ref{eqn:IOPD}) boils down to computing the spatial two-point function $\Delta_{j}(E,\bm{L})$.  As shown in Appendix \ref{app:CD}, the integral (\ref{eqn:1PIES}) can be performed exactly, with final result (\ref{eqn:AEPR})
\begin{equation}
	\Delta_{j}(E,L) = \frac{i}{4\pi L}\exp\Bigg\{\frac{i}{\sqrt{2}}\bigg[\sqrt{R_j^2 + A_j^2} + R_j\bigg]^{1/2}L - \frac{1}{\sqrt{2}}\bigg[\sqrt{R_j^2 + A_j^2} - R_j\bigg]^{1/2}L\Bigg\}~.\label{eqn:EPR}
\end{equation}
in which
\begin{align}	
	R_j & \equiv E^2 - m_j^2~,\notag\\
	A_j & \equiv m_j\Gamma_j~. \label{eqn:DAR}
\end{align}
Note that the exact result in Eq. (\ref{eqn:EPR}) is independent of the orientation of $\bm{L}$~. 

\subsection{Exact Probabilities}
We may now compute the exact oscillation probability via application of Eqs. (\ref{eqn:OPD}), (\ref{eqn:CBD}) and (\ref{eqn:EPR}), and the exact integrated oscillation probability via Eqs. (\ref{eqn:IOPD}), (\ref{eqn:CBD}) and (\ref{eqn:EPR}). It is convenient to define the wavenumber and characteristic inverse decay lengths
\begin{align}
	\omega_j & \equiv \frac{1}{\sqrt{2}}\bigg[\sqrt{R_j^2 + A_j^2} + R_j\bigg]^{1/2}~,\notag\\
	\zeta_j & \equiv \frac{1}{\sqrt{2}}\bigg[\sqrt{R_j^2 + A_j^2} - R_j\bigg]^{1/2}~,\label{eqn:DOZ}
\end{align}
along with 
\begin{equation}
	\label{eqn:DDOZ}
	\Delta \omega_{jk}  \equiv \omega_j - \omega_k~, \qquad \Delta \zeta_{jk}  \equiv \zeta_j - \zeta_k~, \qquad \bar{\zeta}_{jk}  \equiv \zeta_j + \zeta_k~.
\end{equation}
We call $\Delta \omega_{jk}$ the oscillation wavenumber.

Exploiting the unitarity of $U$,  one finds the exact oscillation probability 
\begin{align}	
	P_{\alpha\to\beta}(E,L) 
	& = \Bigg\{\underset{j}{\sum} |U^{\alpha j}|^2|U^{\beta j}|^2e^{-2\zeta_{j}L} + 2\underset{j<k}{\sum} \mbox{Re} \Big[U^{\alpha j}U^{\beta j*}U^{\alpha k *}U^{\beta k}
e^{i\Delta\omega_{jk}L}e^{-\bar{\zeta}_{jk}L}\Big]\Bigg\}\notag\\
	& \quad \times \bigg[\underset{j}{\sum} |U^{\alpha j}|^2e^{-2\zeta_{j}L}\bigg]^{-1}~.\label{eqn:GOPF}
\end{align}
For $\zeta_j \to 0^+$, this has the exact form of the Pontecorvo oscillation formula \cite{Pontecorvo:1968,PDG:2010}. We will show below that within a certain parameter regime, $\Delta \omega_{jk} \simeq (m_k^2 -m_j^2)/2E$, recovering the usual result. The exact integrated oscillation probability is similarly
\begin{align}
	P_{\alpha\to\beta}^{\textrm{I}}(E) 
	& = \Bigg\{\underset{j}{\sum}|U^{\alpha j}|^2|U^{\beta j}|^2/2\zeta_j + 2\underset{j<k}{\sum} \mbox{Re}\bigg[U^{\alpha j}U^{\beta j*}U^{\alpha k *}U^{\beta k}\frac{\bar{\zeta}_{jk} + i\Delta\omega_{jk}}{\bar{\zeta}_{jk}^2 + \Delta\omega_{jk}^2}\bigg]\Bigg\}\notag\\
	& \quad \times \bigg[\underset{j}{\sum} |U^{\alpha j}|^2/2\zeta_j\bigg]^{-1}~.	\label{eqn:GIOPF}
\end{align}
(The $\Delta_j$ normalization $i/4\pi L$ plays an important role in computing the integrals in Eq. (\ref{eqn:IOPD}).) Note that for both Eqs. (\ref{eqn:GOPF}) and (\ref{eqn:GIOPF}) we have not assumed CP conservation. 

\subsection{Two-Flavor Formulae}
It is particularly illuminating to present the oscillation probability and integrated oscillation probability for the case that there are just two flavors. In this case we can choose $U$ to be real, orthogonal: The only physical parameter is the mixing angle and there is no CP violation. We adopt the convention for two flavors that $\alpha = +,-$ and $j = 1,2$. One obtains oscillation probability
\begin{align}
	P_{\alpha \to \beta}(E,L) 
	& = \bigg\{|U^{\alpha 1}|^2|U^{\beta 1}|^2\bigg\}\bigg[|U^{\alpha 1}|^2 + |U^{\alpha 2}|^2e^{-2\Delta \zeta_{21} L}\bigg]^{-1}\notag\\
	&  + \bigg\{|U^{\alpha 2}|^2|U^{\beta 2}|^2\bigg\}\bigg[|U^{\alpha 1}|^2e^{+2\Delta \zeta_{21} L} + |U^{\alpha 2}|^2\bigg]^{-1}\notag\\
	& + \bigg\{2U^{\alpha 1} U^{\beta 1 } U^{\alpha 2 } U^{\beta 2} \cos(\Delta\omega_{12} L)\bigg\}\bigg[|U^{\alpha 1}|^2e^{+\Delta \zeta_{21} L} + |U^{\alpha 2}|^2e^{-\Delta \zeta_{21} L}\bigg]^{-1}~. \label{eqn:ODPE}
\end{align}
Assuming without loss of generality that $\Delta \zeta_{21} >0$, then for $\Delta \zeta_{21}L \gg 1$, the first term is asymptotically constant, the second decays to zero while the third term produces a damped oscillation decaying to zero, with oscillation wavenumber $\Delta \omega_{12}$. In particular, for $\Delta \zeta_{21}L \gg 1$,
\begin{equation}
	\label{eqn:LDUPO}
	P_{\alpha \to \beta}(E,L) \simeq |U^{\beta 1}|^2~.
\end{equation}
If we adopt the notation that $(\phi^\beta)^\dagger$ creates $|\beta\rangle$ and $(\phi^j)^\dagger$ creates $|j\rangle$, then the right side of Eq. (\ref{eqn:LDUPO}) is nothing but $|\langle \beta|1\rangle|^2$. This is the probability of the $\beta$ flavor state being measured as the $j=1$ 1PI state, which has the longer decay length $1/\zeta_1 > 1/\zeta_2$. This behavior is familiar to that found in the $K$ neutral meson system: Since the $K_L$ eigenstate has a much longer decay length than the $K_S$, then the $K_L$ will be exponentially more abundant at large distances from the source compared to the $K_S$ state. As a result, at large distances there is no more oscillation and the oscillation probabilities $K \to K$ or $\overline{K} \to K$ both collapse to $|\langle K|K_L\rangle|^2$. This is exactly the behavior in Eq. (\ref{eqn:LDUPO}).

Before presenting the two-flavor integrated oscillation probability, for convenience we first define
\begin{equation}
	\label{eqn:DXY}
	x \equiv \frac{\Delta\omega_{12}}{\bar{\zeta}_{12}}~,\qquad y \equiv \frac{\Delta\zeta_{21}}{\bar{\zeta}_{12}}~.
\end{equation}
In Sec. \ref{sec:SMS} below we shall verify that $x$ and $y$ reduce to their usual definitions $x \simeq \Delta m /\bar{\Gamma}$ and $y \simeq \Delta \Gamma/2 \bar{\Gamma}$ within a certain regime of the parameters $E$, $m_{1,2}$ and $\Gamma_{1,2}$. With the definitions (\ref{eqn:DXY}), Eq. (\ref{eqn:GIOPF}) reduces to
\begin{align}
	P_{\alpha\to\beta}^{\textrm{I}}(E) 
	& = \Bigg\{|U^{\alpha 1}|^2|U^{\beta 1}|^2(1+y) + |U^{\alpha 2}|^2|U^{\beta 2}|^2(1-y) + 2U^{\alpha 1}U^{\beta 1}U^{\alpha 2}U^{\beta 2}\bigg[\frac{1-y^2}{1+x^2}\bigg]\Bigg\}\notag\\
	& \quad \times \bigg[1 + y\big(|U^{\alpha 1}|^2 - |U^{\alpha 2}|^2\big)\bigg]^{-1}~.\label{eqn:TFIOP}
\end{align}

\subsection{CP Violation and Non-Unitary Diagonalization for Two Flavors}
\label{sec:NUM}
The above oscillation probabilities (\ref{eqn:GOPF}) and (\ref{eqn:GIOPF}) (or (\ref{eqn:OPD}) and (\ref{eqn:IOPD})) may be generalized to the case that $U$ is constant and non-unitary, which is applicable to the study of CP violation in two-flavor neutral meson mixing. In this context, we identify the flavor fields $(\phi^{\pm})^{\dagger}$ as the creation operators of the CP conjugate states $|P^0\rangle$ and $|\overline{P^0}\rangle$, while the 1PI basis $(\phi^{1,2})^{\dagger}$ create the evolution eigenstates $|P_{L,H}\rangle$ respectively.  Comparing Eq. (\ref{eqn:UT}) with the usual notation for $|P^0\rangle$ and $|\overline{P^0}\rangle$ in terms of $|P_{L,H}\rangle$ (assuming CPT symmetry)
\begin{align}
	|P^0\rangle & = \frac{1}{2p}\big(|P_L\rangle + |P_H\rangle\big)~,\notag\\
	|\overline{P^0}\rangle & = \frac{1}{2q}\big(|P_L\rangle - |P_H\rangle\big)~,\label{eqn:POPLH}
\end{align}
then leads to the identification
\begin{equation}
	\label{eqn:UCP}
	U = \frac{1}{2pq}\bordermatrix{ & \mbox{\tiny{1}} & \mbox{\tiny{2}} \cr \mbox{\tiny{$+$}} & q & q\cr \mbox{\tiny{$-$}} & p & -p\cr}~,
\end{equation}
which is non-unitary for $|p/q| \not = 1$. As $|p/q| \not=1$ is sufficient for CP violation in Eq. (\ref{eqn:POPLH}), the consequence of the identification (\ref{eqn:UCP}) is that CP violation in two-flavor mixing is manifested as non-unitary diagonalization of the two-point function. Note also that if $U$ is non-unitary then the 1PI states are no longer orthogonal, as expected for the evolution eigenstates $|P_{L,H}\rangle$. That is $\langle T \phi^{i}(x)\phi^{j\dagger}(y) \rangle \not= \delta_{ij}\Delta_j(x-y)$.  From Eq. (\ref{eqn:GDU}) we have for non-unitary $U$
\begin{equation}
	\label{eqn:CBDNU}
	\Delta_{\alpha\beta}(E,L) = U^{\alpha j}(U^{-1})^{j\beta}\Delta_{j}(E,L) ~,
\end{equation}
and from Eqs. (\ref{eqn:EPR}) and (\ref{eqn:DOZ}) one then finds exact amplitudes
\begin{align}
	 \big|\Delta_{\pm\pm}(E,L)\big|^2 & = \frac{1}{32\pi^2 L^2}e^{-\bar{\zeta}_{12}L}\Big[\cosh(\Delta\zeta_{12}L) + \cos(\Delta \omega_{12} L)\Big]~,\notag\\
	\big|\Delta_{+-}(E,L)\big|^2 & = \frac{|q|^2}{|p|^2}\frac{1}{32\pi^2 L^2}e^{-\bar{\zeta}_{12}L}\Big[\cosh(\Delta\zeta_{12}L) - \cos(\Delta \omega_{12} L)\Big] \notag\\
	& = \frac{|q|^4}{|p|^4}\big|\Delta_{-+}(E,L)\big|^2 ~.\label{eqn:ECPA}
\end{align}
These strongly resemble the amplitudes found from the usual meson mixing quantum mechanical analysis (see below), except that here a spatial dependence has replaced the usual time dependence and $\Delta\omega_{jk}$, $\Delta\zeta_{jk}$ and $\bar{\zeta}_{jk}$ have the form presented in Eq. (\ref{eqn:DDOZ}). One finds exact integrated oscillation probabilities
\begin{align}
	P^{\textrm{I}}_{+\to +}  & = \frac{2 + x^2 - y^2}{2 + x^2 - y^2 + \big|q/p\big|^2(x^2 + y^2)}~,\notag\\
	P^{\textrm{I}}_{+\to -} & = \frac{x^2 + y^2}{x^2 + y^2 + \big|p/q\big|^2(2 + x^2 - y^2)}~,\label{eqn:EIPPQ}
\end{align}
and one can also contemplate measuring the ratio of the amplitudes for oscillation into either flavor state
\begin{align}
	F  &\equiv \frac{\mathcal{A}^{\textrm{I}}_{+-}(E)}{\mathcal{A}^{\textrm{I}}_{++}(E)} = \frac{\int{\rm 
	d}L|\Delta_{+-}(E,L)|^2}{\int{\rm	d}L|\Delta_{++}(E,L)|^2} \notag\\
	& = \bigg|\frac{q}{p}\bigg|^2\frac{x^2 + y^2}{2 + x^2 - y^2}~.\label{eqn:EARPQ}
\end{align}

Let us compare the exact results in Eqs. (\ref{eqn:EIPPQ}) and (\ref{eqn:EARPQ}) with the analogous formulae obtained via the usual quantum mechanical treatment of neutral meson oscillations. Following Refs. \cite{Branco:1999cp,Bigi:2009cp} we can write down the time evolution for an initial pure $|P^0\rangle$ and $|\overline{P^0}\rangle$ state
\begin{align}
	|P^0(t)\rangle=g_+(t)|P^0\rangle+\frac{q}{p}~g_-(t)|\overline{P^0}\rangle\notag\\
	|\overline{P^0}(t)\rangle=\frac{p}{q}~g_-(t)|P^0\rangle+g_+(t)|\overline{P^0}\rangle~,\label{eqn:QM}
\end{align}
where
\begin{equation}
	|g_\pm(t)|^2=\frac{e^{-\bar{\Gamma} t}}{2}\Big[\cosh(\Delta\Gamma t/2)\pm\cos(\Delta m t)\Big]~.\label{eqn:EVO}
\end{equation}
In the standard notation 
\begin{equation}
	x=\Delta m/\bar{\Gamma}~,\qquad y=\Delta\Gamma/2\bar{\Gamma}~, 
\end{equation}
where $\bar{\Gamma}=(\Gamma_1+\Gamma_2)/2$, the two formulae in Eqs. (\ref{eqn:EIPPQ}) should be compared with
\begin{equation}
	\frac{\int_0^\infty{\rm d}t~|\langle P^0|P^0(t)\rangle|^2}{\int_0^\infty{\rm d}t~\Big[|\langle P^0|
	P^0(t)\rangle|^2+|\langle \overline{P^0}| P^0(t)\rangle|^2\Big]}=\frac{2+x^2-y^2}{2+x^2-y^2+|q/
	p|^2(x^2+y^2)}
\end{equation}
and
\begin{equation}
	\frac{\int_0^\infty{\rm d}t~|\langle \overline{P^0}|P^0(t)\rangle|^2}{\int_0^\infty{\rm 
	d}t~\Big[|\langle P^0|P^0(t)\rangle|^2+|\langle \overline{P^0}| P^0(t)\rangle|^2\Big]}=
	\frac{x^2+y^2}{x^2+y^2+|p/q|^2(2+x^2-y^2)}~.
\end{equation}
 Finally Eq. (\ref{eqn:EARPQ}) should be compared to
\begin{equation}
	\frac{\int_0^\infty{\rm d}t~|\langle \overline{P^0}|P^0(t)\rangle|^2}{\int_0^\infty{\rm 
	d}t~|\langle P^0|P^0(t)\rangle|^2}=\left|\frac{q}{p}\right|^2\frac{x^2+y^2}{2+x^2-y^2}~.
\end{equation}
Our exact results are in perfect agreement with those of the usual analysis, except that in Eqs. (\ref{eqn:EIPPQ}) and (\ref{eqn:EARPQ}), the parameters $x$ and $y$ have the more general definitions encoded in Eqs. (\ref{eqn:DAR}), (\ref{eqn:DOZ}) and (\ref{eqn:DXY}). Once again, in Sec. \ref{sec:SMS} below we shall verify that $x$ and $y$ reduce to their usual definitions $x \simeq \Delta m /\bar{\Gamma}$ and $y \simeq \Delta \Gamma/2 \bar{\Gamma}$ within a certain regime of the parameters $E$, $m_{1,2}$ and $\Gamma_{1,2}$. 

Before proceeding, please note that the oscillation probabilities (\ref{eqn:GOPF}), (\ref{eqn:GIOPF}), (\ref{eqn:ODPE}), (\ref{eqn:TFIOP}) and (\ref{eqn:ECPA})  presented in this section are a function of only $\Delta\omega_{jk}$, $\Delta\zeta_{jk}$, $\bar{\zeta}_{jk}$ or $\zeta_j$ (of the latter three variables, only two are independent). Consequently, specifying just $\omega_j$ and $\zeta_j$ is sufficient to specify the oscillation probabilities. This shall be our practice throughout the remainder of this paper.

\section{Regimes}
\label{sec:R}

The results presented in Sec. \ref{sec:EOP} are elegant and concise, but their physical interpretation is not obvious. However, in various regimes of the parameters $E$, $m_j$ and $\Gamma_j$, our exact results for the wavenumber and characteristic inverse decay lengths $\omega$ and $\zeta$ reduce to simpler expressions with clear physical meanings. In this section we explore several different regimes of physical interest, and show that in certain regimes our results reproduce the well-known neutrino and neutral meson oscillation formulae.

\subsection{Particle Regime}
\label{sec:PR}
The first regime of interest is the case
\begin{equation}
	\label{eqn:DPR}
	R_j \gg A_j~,~~\mbox{i.e.}~~E^2 -m_j^2 \gg m_j\Gamma_j~.
\end{equation}
It is straightforward to expand Eq. (\ref{eqn:DOZ}) about $A_j = 0$, and to leading order in $A_j/R_j$ one obtains the wavenumber and characteristic inverse oscillation lengths
\begin{equation}
	\label{eqn:PROZ}
	\omega_j  \simeq \sqrt{E^2 -m_j^2}~,\qquad\zeta_j  \simeq \frac{m_j\Gamma_j}{2\sqrt{E^2 - m_j^2}}~.
\end{equation}
The oscillation probabilities (\ref{eqn:GOPF}) and (\ref{eqn:GIOPF}) follow immediately from this and Eqs. (\ref{eqn:DDOZ}), as do their two-flavor versions (\ref{eqn:ODPE}) and (\ref{eqn:TFIOP}). In particular, in this regime we have to leading order in $A_j/R_j$
\begin{equation} 
	\label{eqn:PRXY}
	x  \simeq 2\frac{\sqrt{R_1} - \sqrt{R_2}}{A_1/\sqrt{R_1} + A_2/\sqrt{R_2}}~,\qquad y  \simeq \frac{A_1/\sqrt{R_1} - A_2/\sqrt{R_2}}{A_1/\sqrt{R_1} + A_2/\sqrt{R_2}}~. 
\end{equation}

In this regime, the spatial two-point function for a 1PI state
\begin{equation}
	\label{eqn:PRSP}
	\Delta_j(E,L) \simeq \frac{i}{4\pi L} \exp\Bigg\{i\sqrt{E^2 -m_j^2}L - \frac{m_j\Gamma_j}{2\sqrt{E^2-m_j^2}}L\Bigg\}~.
\end{equation}
For $\Gamma_j \to 0^+$, this looks precisely like the propagator of an on-shell one-particle state with momentum $p = (E^2 -m_j^2)^{1/2}$. We therefore call the regime (\ref{eqn:DPR}) the particle regime. The resemblance of the amplitude (\ref{eqn:PRSP}) to that of a particle suggests that we should obtain both the neutrino and meson mixing oscillation formula within the particle regime. We explicitly verify this in the next two sections. 

The physical meaning of the spatial two-point function perhaps becomes more clear if we define analogous Lorentz factors and proper time
\begin{equation}
	\label{eqn:DBPR}
	\gamma_j  = E/m_j~, \qquad \beta_j  = \sqrt{1 - m_j^2/E^2}~,\qquad\tau_j  = L/\gamma_j\beta_j~.
\end{equation}
Substituting these into Eq. (\ref{eqn:PRSP}) we obtain the spatial two-point function in terms of $\tau_j$ instead of $L$, which we can interpret as a `rest frame' propagator. Explicitly, 
\begin{equation}
	\Delta_j(\tau_j) \simeq \frac{i}{4\pi L}\exp\bigg\{i m_j(\gamma^2_j-1)\tau_j - \frac{\Gamma_j\tau_j}{2}\bigg\}~.
\end{equation}
The second term in the exponential looks like the usual rest frame decay of an unstable particle, and in particular it is clear that $\Gamma_j$ can be interpreted as the rest frame decay rate. The first term looks like the usual proper time evolution of a particle, except for the $\gamma^2-1$ factor. This factor arises because $pL$ is not a Lorentz invariant, but rather $E\gamma\tau - pL = m\tau$ is. It is a consequence of the experiment measure $E$ rather than the time of transit between the source and detector.

Let us now proceed to verify that the usual neutrino and neutral meson mixing oscillation formulae are obtained in this particle regime. 

\subsection{Small Mass Splitting: Neutral Meson Oscillation}
\label{sec:SMS}
In all known neutral meson systems, the mass difference between the two mass eigenstates is extremely small in comparison with their masses. For the $K$, $D$, $B_d$ and $B_s$ neutral meson systems one finds \cite{PDG:2010}
\begin{equation}
	\label{eqn:MSD}
	\bigg(\frac{\Delta m}{m}\bigg)_K  \sim \bigg(\frac{\Delta m}{m}\bigg)_D \sim10^{-14}~,\qquad \bigg(\frac{\Delta m}{m}\bigg)_{B_d}  \sim 10^{-13}~,\qquad \bigg(\frac{\Delta m}{m}\bigg)_{B_s}  \sim 10^{-12}~.
\end{equation}
It seems then, that the appropriate regime for neutral meson oscillation is the particle regime with the additional constraint that the mass splitting is small. We define the mean mass $m$ and mass splitting $\Delta m$ via $m_1 = m + \Delta m/2$ and $m_2 = m - \Delta m/2$, so the small mass splitting limit is $\Delta m/m \ll 1$. We also define $y_0 \equiv \Delta \Gamma/2\bar{\Gamma}$, in which $\bar{\Gamma} \equiv (\Gamma_1 + \Gamma_2)/2$ and $\Delta \Gamma \equiv \Gamma_2 - \Gamma_1$. 

Expanding the particle regime expressions (\ref{eqn:PRXY}) for $x$ and $y$ in the small mass splitting limit is complicated by the fact that neither $x$ nor $y$ can be expressed as function of $\Delta m/m$ alone. However, one may show that 
\begin{align}
	x & \simeq \frac{\Delta m}{\bar{\Gamma}}\bigg[1 + \frac{y_0}{2\beta^2}\frac{\Delta m}{m} + \sum_{p=2}^\infty \frac{X_p(y_0,m,E)}{2^p\beta^{2p}}\bigg(\frac{\Delta m}{m}\bigg)^p\bigg]\notag\\
	y & \simeq \frac{\Delta \Gamma}{2\bar{\Gamma}}\bigg[1 + \frac{1- y_0^2}{2\beta^2y_0}\frac{\Delta m}{m} + \sum_{p=2}^\infty \frac{Y_p(y_0,m,E)}{2^p\beta^{2p}}\bigg(\frac{\Delta m}{m}\bigg)^p\bigg]~,\label{eqn:XYSM}
\end{align}
for which $X_p$ and $Y_p$ are rational functions of $y_0$, $m$ and $E$. The parameter $\beta$ is defined as in Eqs. (\ref{eqn:DBPR}), but for mass $m$. In general, any of $X_p$, $Y_p$ or $1/\beta$ could be arbitrarily large for some configuration of the parameters $y_0$, $m$ and $E$, so the expansions (\ref{eqn:XYSM}) are not always well-controlled power series in $\Delta m/m$. However, it's plausible that the expansions are well-controlled in the parameter space regimes, and respective very small mass splittings (\ref{eqn:MSD}), relevant to the neutral meson systems. In such regimes, we then obtain for sufficiently small mass splittings
\begin{equation}
	x \simeq \frac{\Delta m}{\bar{\Gamma}}~,\qquad y \simeq \frac{\Delta \Gamma}{2\bar{\Gamma}}~.
\end{equation}
These are precisely the usual definitions for the parameters $x$ and $y$ in the neutral meson mixing formalism. Moreover, in terms of the `proper time' $\tau$ - defined for $m$ in Eqs. (\ref{eqn:DBPR}) - the same expansion in small mass splitting renders the CP violating amplitudes (\ref{eqn:ECPA})
\begin{align}
	\big|\Delta_{++}(E,L)\big|^2 &\simeq \frac{1}{32\pi^2 L^2}\exp(-\bar{\Gamma}\tau)\Big[\cosh(\Delta\Gamma \tau/2) + \cos(\Delta m \tau)\Big]~,\notag\\
	\big|\Delta_{+-}(E,L)\big|^2 & \simeq \frac{1}{32\pi^2 L^2}\frac{|q|^2}{|p|^2}\exp(-\bar{\Gamma}\tau)\Big[\cosh(\Delta\Gamma \tau/2) - \cos(\Delta m \tau)\Big]~.
\end{align}
Up to a normalization of  $1/16\pi^2L^2$, these are exactly the time evolution amplitudes found within the usual meson mixing analysis \cite{Branco:1999cp,Bigi:2009cp} as is evident in Eq. (\ref{eqn:EVO}). The physical interpretation of $\tau$ is the proper time elapsed in the rest frame of a classical particle with mass $m$ and lab frame energy $E$ that traverses a distance $L$. 

Since we have already verified that, in terms of $x$ and $y$, our integrated oscillation probabilities match those found in the usual treatment, we have therefore recovered the well-known meson mixing amplitudes and time-integrated probabilities from the structure of the two-point function alone. Our analysis, however, also implies that the usual meson mixing results are valid \emph{only} within the small mass splitting particle regime. Outside this regime the more general results of Sec. \ref{sec:NUM} will apply.

An immediate question is whether the regimes of validity of our derivation and the standard quantum mechanical one disagree. The standard derivation is performed in terms of time evolution, and requires a \emph{common} proper time for the mass eigenstates \cite{Branco:1999cp,Bigi:2009cp}, which are one-particle states. If the energy of the oscillation experiment is fixed, as we assume throughout this paper, then this assumption is equivalent to assuming $\Delta m \to 0$. So the standard derivation is similarly applicable only in the small mass splitting limit.

\subsection{(Ultrarelativistic) Stable Particle Regime: Neutrino Oscillation}
For a neutrino oscillation experiment  we expect the neutrinos to be ultrarelativistic and stable in the lab frame. The ultrarelativistic stable limit of the particle regime corresponds to $E \gg m_j$ and $\Gamma_j \to 0^+$ for all $j$. Expanding in this limit, the wavenumbers and characteristic inverse decay lengths (\ref{eqn:PROZ}) become, to leading order in $m_j/E$,
\begin{equation}
	\label{eqn:SPROZ}
	\omega_j \simeq E - \frac{m_j^2}{2E}~,\qquad \zeta_j = 0~,
\end{equation}
so that
\begin{equation}
	\label{eqn:SPR}
	\Delta_j(E,L) \simeq \frac{i}{4\pi L}\exp\Bigg\{iEL - i\frac{m_j^2}{2E}L\Bigg\}~.
\end{equation} 
Applying Eqs. (\ref{eqn:GOPF}) and the unitarity of $U$ leads immediately to 
\begin{align}
	P_{\alpha\to\beta}(E,L) 
	& \simeq \sum_j|U^{\alpha j}|^2|U^{\beta j}|^2 + 2\sum_{j<k} \mbox{Re} \bigg[ U^{\alpha j} U^{\beta j *} U^{\alpha k *} U^{\beta k} \exp\bigg\{ -i\frac{\Delta m_{jk}^2}{2E}L\bigg\}\bigg]\notag\\
	& = \delta_{\alpha\beta} + 2\sum_{j<k} \mbox{Im} \Big[ U^{\alpha j} U^{\beta j *} U^{\alpha k *} U^{\beta k}\Big]\sin\bigg(\frac{\Delta m_{jk}^2}{2E}L\bigg)\notag\\
	&\quad - 4\sum_{j<k} \mbox{Re} \Big[ U^{\alpha j} U^{\beta j *} U^{\alpha k *} U^{\beta k}\Big]\sin^2\bigg(\frac{\Delta m_{jk}^2}{4E}L\bigg)~,\label{eqn:NOF}
\end{align}
where $\Delta m_{jk}^2 \equiv m_j^2 - m_k^2$. This is precisely the Pontecorvo neutrino oscillation formula. Hence we have derived the neutrino oscillation formula in a purely quantum field theoretic formalism, involving just the structure of the spatial two-point function. Comparing Eqs. (\ref{eqn:PROZ}) and (\ref{eqn:SPROZ}), it is straightforward to generalize this result to just the stable particle regime $E > m_j$, $\Gamma_j \to 0^+$ via the replacement in Eq. (\ref{eqn:NOF})
\begin{equation}
-\frac{\Delta m_{jk}^2}{2E} \rightarrow \sqrt{E^2 -m_j^2} - \sqrt{E^2 - m_k^2}~.
\end{equation}

\subsection{(Deep) Virtual Regime}
Having verified that our exacts results reduce to the expected results for both the neutral meson and neutrino systems, let us now exploit the generality of  Eqs. (\ref{eqn:EPR}), (\ref{eqn:DOZ}) and (\ref{eqn:GOPF}) to push $E$, $m_j$ and $\Gamma_j$ into non-standard, though physically relevant, regimes of parameter space. So far we have only considered regimes for which $E>m_j$, so let us now consider the case
\begin{equation}
	-R_j \gg A_j~,~~\mbox{i.e.}~~m_j^2 - E^2 \gg m_j\Gamma_j~,
\end{equation}
which we call the virtual regime for reasons outlined below. If also $E \ll m_j$, then we call this the deep virtual regime. As we will investigate in Sec. \ref{sec:MR}
below, the virtual regime is particularly interesting if one 1PI state is very heavy compared to another.

In the virtual regime, the wavenumber and characteristic inverse decay lengths become to leading order in $A_j/|R_j|$
\begin{equation}
	\label{eqn:DTROZ}
	\omega_j \simeq \frac{m_j\Gamma_j}{2\sqrt{m_j^2 - E^2}}~,\qquad \zeta_j  \simeq \sqrt{m_j^2 - E^2}~.
\end{equation}
Again, the oscillation probabilities (\ref{eqn:GOPF}) and (\ref{eqn:GIOPF}) follow immediately from this and Eqs. (\ref{eqn:DDOZ}). Within this regime, two flavors with a sufficiently small mass splitting have integrated oscillation probability described by the parameters
\begin{equation}
	x \simeq \frac{\Delta \Gamma}{2\bar{\Gamma}}~,\qquad y \simeq \frac{\Delta m}{\bar{\Gamma}}~.
\end{equation}
These are, of course, just a swap of the usual parameters one sees in the particle regime. 

In the virtual regime Eq. (\ref{eqn:EPR}) becomes
\begin{equation}
	\label{eqn:EXP}
	\Delta_j(E,L) \simeq \frac{i}{4\pi L}\exp\Bigg\{i\frac{m_j\Gamma_j}{2\sqrt{m_j^2-E^2}}L - \sqrt{m_j^2-E^2}L\Bigg\}~.
\end{equation}
The spatial two-point function no longer looks like that of a one-particle state. This is especially clear in the stable virtual case $m_j> E$ and $\Gamma_j \to 0 ^+$, for which $\Delta_j(E,L)$ is just an exponential decay: This was noticed previously in Ref. \cite{Okun:1982sr}. Note also that for the unstable case, the wavenumber is determined by the decay rate, rather than by a momentum $(E^2 -m_j^2)^{1/2}$, and vice versa for the characteristic inverse decay length.

Let us briefly comment on the physical interpretations of the virtual regime results, from which we derive its name. As explained in Sec. \ref{sec:EA}, the spatial two-point function $\Delta_j(E,L)$ does not encode the propagation of just a single particle with a definite momentum. Rather, as suggested by Eq. (\ref{eqn:1PIES}), we may think of the spatial two-point function as the continuous sum of a set of propagators, each corresponding to the propagation of a momentum eigenstate. The condition $E < m_j$ then implies all these momentum eigenstates must be off-shell, so in this case $\Delta_j(E,L)$ includes no on-shell propagating particles. That is they are virtual particles, whence the regime name. Alternatively, $E<m_j$ is analogous to the usual quantum mechanical tunnelling condition, with the mass acting as the potential barrier. From either point of view, we emphasize that we should expect $\Delta(E,L)$ to be exponentially suppressed in the stable case, precisely as we see in Eq. (\ref{eqn:EXP}).

\subsection{Mixed Regime}
\label{sec:MR}
It is interesting to consider the case that different 1PI states occupy different regimes. For example, we could consider a two-flavor oscillation in the case that one 1PI state is in the particle regime, while the other is in the virtual regime. We call such a case the mixed regime. This scenario doesn't occur for the neutral meson or neutrino systems because the mass splitting between 1PI states is very small compared to $E$. However, the large mass hierarchy of the quark sector combined with the possibility of quark oscillations \cite{Grossman:2011,Pilaftsis:1997}, provides a natural setting in which we may contemplate a mixed regime oscillation. 

\begin{figure}[t]
  \begin{picture}(261,128) (0,0)
    \SetWidth{0}
    \SetColor{Black}
    \Line[dash,dashsize=4,arrow,arrowpos=0.5,arrowlength=5,arrowwidth=2,arrowinset=0.2](172,95)(217,109)
    \Line[dash,dashsize=4,arrow,arrowpos=0.5,arrowlength=5,arrowwidth=2,arrowinset=0.2](172,86)(215,73)
    \Line[dash,dashsize=4,arrow,arrowpos=0.5,arrowlength=5,arrowwidth=2,arrowinset=0.2](64,82)(222,28)
    \Line[arrow,arrowpos=0.5,arrowlength=5,arrowwidth=2,arrowinset=0.2](68,86)(162,91)
    \Line[arrow,arrowpos=0.5,arrowlength=5,arrowwidth=2,arrowinset=0.2](1,113)(49,89)
    \GOval(221,64)(63,14)(0){0.882}
    \GOval(59,86)(12,12)(0){0.882}
    \GOval(167,90)(12,12)(0){0.882}
    \Text(29,104)[lb]{\Black{$t$}}
    \Text(57,83)[lb]{\Black{S}}
    \Text(164,87)[lb]{\Black{D}}
    \Text(217,58)[lb]{\Black{$X_\beta$}}
    \Text(100,95)[lb]{\Black{$b,b'$}}
  \end{picture}
\caption{Quark oscillation experiment $t \to X_\beta$.}
\label{fig:QE}
\end{figure}
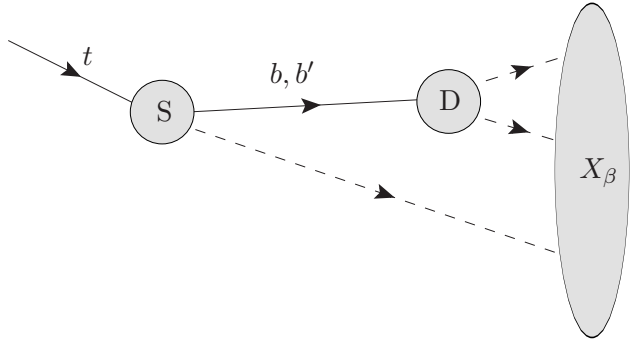

For concreteness, let us suppose there is a fourth quark doublet $(t',b')$, with masses much larger than the top quark mass. The existence of a fourth quark family is strongly constrained by the electroweak weak precision measurements \cite{Hong:2001,Kribs:2007,Hung:2008}, but nonetheless phenomenologically still perfectly viable (see Ref. \cite{PDG:2010} for mass bounds). A quark oscillation experiment could then involve a top quark decaying to a final state $t \to X_\beta$ via intermediate $b$ or $b'$ down-type quarks. The generic diagrammatic form of such an experiment is shown in Fig. \ref{fig:QE}.

Let us adopt the following notation. The flavor of the down-type quarks is determined by their up-type partner, so we denote the down-type flavor quarks by $b_t$ and $b_{t^\prime}$. That is, $\alpha = t,t'$. Correspondingly the 1PI states are denoted $b$ and $b^\prime$, so $j=b,b'$. The idea here is that the top quark produces the flavor quark $b_t$ at the source vertex, S, while the generic final state $X_\beta$ in the detector can tag the flavor at vertex D. In order to describe the physics of this experiment using our formalism, and for simplicity, we also assume the following:
\begin{enumerate}[label=\roman{enumi})]
\item The amplitude of the experiment is described by Eq. (\ref{eqn:DEA}).
\item We neglect the presence of the other two down-type quarks $d$ and $s$, and consider an effective two-flavor mixing between the third and fourth quark generations. Consequently, the final state $X_\beta$ only measures flavors $\beta = t,t'$.
\item The $b$ is in the stable particle regime ($E > m_b$ and $\Gamma_b \to 0^+$), while $b'$ in the stable virtual regime ($E<m_{b'}$ and $\Gamma_{b'} \to 0^+$).
\item The energy, $E$, exchanged between S and D can be precisely measured. 
\item The 2$\times$2 mixing matrix $U$, which diagonalizes the 1PI function, is unitary.
\end{enumerate}
The extent to which these assumptions are applicable to an actual quark oscillation experiment is questionable. Our intent is merely to demonstrate that with such assumptions, we can perhaps gain insight into the physics of quark oscillations by use of our formalism. 

With these assumptions, we have wavenumber and characteristic inverse decay lengths
\begin{equation}
	\omega_b = \sqrt{E^2 - m_b^2}~,\qquad \omega_{b'} = 0~,\qquad \zeta_b =0~,\qquad \zeta _{b'} = \sqrt{m_{b'}^2 - E^2}~,
\end{equation}
so that the oscillation wavenumbers
\begin{equation}
	\Delta\omega_{bb'}  = \sqrt{E^2 - m_b^2}~,\qquad\Delta\zeta_{b'b}  = \sqrt{m_{b'}^2 - E^2}~.
\end{equation}
The oscillation probabilities follow immediately from Eqs. (\ref{eqn:ODPE}) and (\ref{eqn:TFIOP}), while the corresponding spatial two-point functions
\begin{align}
	\Delta_b(E,L) & = \frac{i}{4\pi L}\exp\bigg\{i\sqrt{E^2-m_b^2}L\bigg\}~,\notag\\
	\Delta_{b'}(E,L) & =  \frac{i}{4\pi L}\exp\bigg\{-\sqrt{m^2_{b'} - E^2}L\bigg\}~.
\end{align}
In particular, note that the $b'$ two-point function is exponentially suppressed, as we expect for a virtual particle. Further, the integrated probability has parameters
\begin{equation}
	x = \sqrt{\frac{E^2 - m_b^2}{m_{b'}^2 - E^2}}~,\qquad y = 1~,
\end{equation}
so that in the two-flavor integrated oscillation probability (\ref{eqn:TFIOP}) there is no longer any interference term --- and hence no oscillation --- between the two mass eigenstates. From Eq. (\ref{eqn:GIOPF}) one finds integrated oscillation probability
\begin{equation}
	P_{t\to\beta}^{\textrm{I}}(E) = 2 \frac{|U^{t b}|^2|U^{\beta b}|^2}{1 + |U^{tb}|^2 - |U^{t b'}|^2}~.
\end{equation}
(Note that convergence of the integrated amplitude with $\zeta_b = 0$ is ensured by the usual $i\epsilon$ term, which is taken to zero after integration. Equivalently, since $m_b\Gamma_b$ acts as the $\epsilon$ throughout this paper, the stable limit is determined by taking $\Gamma_b \to 0^+$ after integration over $L$.) As expected, the probability is controlled purely by the mixing of the flavor states with the particle-like 1PI state $b$ which is in the stable particle regime. 

\subsection{Threshold Regime}
One last regime of interest, which to the knowledge of the authors has not been previously discussed, is the case
\begin{equation}
	|R_j| \ll A_j~,~~\mbox{i.e.}~~|E^2 -m_j^2| \ll m_j\Gamma_j~.
\end{equation}
We call this the threshold regime, since $E \simeq m_j$. To zeroth order in $|R_j|/A_j$, the wavenumber and characteristic inverse lengths reduce to 
\begin{equation}
	\omega_j = \zeta_j \simeq \sqrt{\frac{m_j\Gamma_j}{2}}~,
\end{equation}
and the oscillation probabilities follow as usual.
This time (\ref{eqn:EPR}) becomes
\begin{equation}
	\Delta_{j}(E,L) \simeq \frac{i}{4\pi L}\exp\Bigg\{ (i-1)\sqrt{\frac{m_j\Gamma_j}{2}}L\Bigg\}~.
\end{equation}
Here, curiously, the inverse decay length and wavenumber both depend on the geometric mean of the decay rate and mass, and coincide. We are unaware of an intuitive physical reason why they should coincide at threshold. There is, however, a limited particle analog to this behavior. If we were to interpret $\omega_j$ as the momentum, as we did in the particle regime, then we would have
\begin{equation}
	p_j^2\simeq m_j^2-m_j\Gamma_j/2~.
\end{equation}
That is, the 1PI states can be thought of as virtual particles slightly perturbed from the mass shell if $\Gamma_j \ll m_j$.

In the threshold regime, a small mass splitting for two flavors results in 
\begin{equation}
	x = y \simeq \frac{\sqrt{\Gamma_1} - \sqrt{\Gamma_2}}{\sqrt{\Gamma_1} + \sqrt{\Gamma_2}}
\end{equation}
while if also the decay rates have a small splitting,  $\Gamma_{1,2} = \Gamma \mp \Delta \Gamma$, $\Delta \Gamma \ll \Gamma$, then
\begin{equation}
	x = y \simeq -\Delta m/m - \Delta \Gamma/\Gamma \ll 1~.
\end{equation}
A well-motivated example of oscillation in which the threshold regime is applicable to both 1PI states is unknown to the authors. Despite this, we do wish to emphasize that the generality of Eqs. (\ref{eqn:GOPF}) and (\ref{eqn:GIOPF}) permits exploration of parameter regimes in which a quantum mechanical treatment might be unfeasible. 

\section{Conclusions}
In this paper we have used only the structure of the spatial two-point function $\Delta_{\alpha\beta}(E,L)$ to derive general flavor oscillation probability formulae for unstable fields. We have not only shown that this structure reproduces the usual Pontecorvo neutrino oscillation formulae and time-integrated (CP violating) neutral meson mixing formulae, but we have also found generalized exact expressions with natural physical interpretations in several different parameter regimes. Our results for the stable particle and stable virtual regimes agree with the results of Ref. \cite{Okun:1982sr} for stable fermions. However, our exact oscillation probabilities for unstable fields and the analysis of the unstable particle, threshold and virtual regimes has not been previously presented. 

The advantages of the formalism we have employed in this paper are several. The exact computability and integrability of $\Delta_{\alpha\beta}(E,L)$ permitted us to obtain exact, elegant probability oscillation formulae. Moreover, the choice of reference frame throughout this paper is the unambiguous laboratory frame: There is no need in our approach to contemplate mass eigenstate rest frames and proper times. To the extent that complicating effects such as coherence, finite detector and source size, non-trivial source and detector physics, and measurement uncertainty can be neglected, our results provide an instructive leading order description of the physics of flavor oscillation, that is valid over the entire $E,m,\Gamma$ parameter space.

In terms of future work, keeping in mind the large existing Literature on this subject, perhaps the most interesting avenue left to explore is the analogous formalism for flavor oscillation in matter, that is, the Mikheyev-Smirnov-Wolfenstein effect.

\acknowledgments
The authors thank Yuval Grossman, Jo\~ ao P. Silva and Philip Tanedo for helpful discussions. This work is supported by NSF grant number
PHY-0757868.

\appendix

\section{Diagonalization of the exact propagator}\label{sec:AB}

In this appendix we discuss the subtleties involved in diagonalizing the exact propagator in Eq. (\ref{eqn:EP}),
\begin{equation}
      \Delta_{\alpha\beta}(p^2) = \bigg[\frac{i}{p^2\bm{1} - M^2(p^2)}\bigg]_{\alpha\beta}~.
\end{equation}

In general the exact propagator $\Delta(p^2)$ is not necessarily Hermitian. It is therefore not always diagonalizable by a unitary matrix and may not even be diagonalizable at all. If, however, $\Delta(p^2)$ \emph{is} diagonalizable by some invertible matrix $U$, then observe that: We should generally expect  $U$ to be a function of $p^2$, $U = U(p^2)$, since $\Delta = \Delta(p^2)$; The diagonalizability of $\Delta_{\alpha\beta}(p^2)$ is equivalent to that of $M_{\alpha\beta}^2(p^2)$, since if one is diagonalizable by $U(p^2)$ then so is the other. 

Keeping these two observations in mind, in the standard field theoretic oscillation formalism one first diagonalizes the tree-level Lagrangian mass terms, thus obtaining free propagators for the mass eigenstates, and then one can construct two-point amplitudes perturbatively. For example, the well-known PMNS (CKM) matrix diagonalizes the lepton (quark) masses in the Standard Model with right-handed neutrinos (SM + $\nu_R$). However, in general such a diagonalization does not persist to all orders in perturbation theory. In particular, in the SM + $\nu_R$ model the flavor changing 1PI functions are zero at tree-level, but receive non-zero contributions at loop level, which are small due to the GIM mechanism. To see this, note that the exact propagator for a left-handed neutrino in the mass basis is
\begin{equation}
\Delta_{ij}(\slashed{p})= \frac{i\delta_{ij}}{\slashed{p} - m_j}+ 
\begin{minipage}{0.33\linewidth}
\begin{center}
  \begin{picture}(130,60) (480,-140)
    \SetWidth{1.0}
    \SetColor{Black}
    \Line[arrow,arrowpos=0.5,arrowlength=6.5,arrowwidth=2.5,arrowinset=0.2](481,-115)(605,-115)
    \SetWidth{1.0}
    \PhotonArc[clock](542,-121.643)(34.643,168.945,11.055){3}{6.5}
    \Text(552,-84)[lb]{\normalsize{\Black{$W^+$}}}
    \Text(504,-130)[lb]{\normalsize{\Black{$U^{i\alpha}_{\textrm{\tiny{PMNS}}}$}}}
    \Text(572,-130)[lb]{\normalsize{\Black{$U^{j\alpha*}_{\textrm{\tiny{PMNS}}}$}}}
    \Text(542,-129)[lb]{\normalsize{\Black{$\ell^\alpha$}}}
    \Text(476,-126)[lb]{\normalsize{\Black{$\nu^i$}}}
    \Text(608,-126)[lb]{\normalsize{\Black{$\nu^j$}}}
  \end{picture}
\end{center}
\end{minipage}
+ \ldots~,
\end{equation}
in which we use the usual SM notation. The mass splittings of the leptons ensure that the neutrino exact propagator is not diagonal at all loop orders in the PMNS basis. Another manifestation of this effect is that flavor changing neutral currents do not appear at tree level, but they do appear at higher loop orders. The moral is that if the propagator is diagonalizable in the exact theory, it is generally diagonalizable by a $p^2$-dependent matrix, which is different from the PMNS or CKM matrix at subleading order and not necessarily unitary.

Due to the GIM suppression, it is common in oscillation formalisms to neglect this effect because it occurs at subleading order in perturbation theory. Instead, one presumes that the propagator is diagonalized by the constant, unitary PMNS or CKM matrix. In this paper we make a similar assumption in Eq. (\ref{eqn:GDU}) in the main text. The validity of this assumption is model dependent, and a discussion of the general circumstances under which it applies is beyond the scope of this paper. Nonetheless, as the above SM + $\nu_R$ example demonstrates, it is true at leading order in perturbation theory for certain important theories.

We emphasize finally that rather than arising from a diagonalization of the bare classical Lagrangian, the matrix $U$ here diagonalizes the exact propagator $\Delta_{\alpha\beta}(p^2)$, which includes all quantum corrections. For the SM + $\nu_R$ example discussed above, $U$ therefore coincides at zeroth order with the neutrino PMNS or quark CKM matrix. However, in general $U$ acts as the mixing matrix between the flavor field basis and the 1PI basis (defined in the main text), rather than between the flavor basis and the mass basis of the classical Lagrangian.  

\section{Computation of $\Delta_{j}(E,\bm{L})$}
\label{app:CD}
In this appendix we compute the integral in Eq. (\ref{eqn:1PIES}): 
\begin{equation}
	\label{eqn:A1PIES}	
	\Delta_j(E,\bm{L}) = \int \frac{d^3p}{(2\pi)^3} \frac{ie^{i\bm{p}\cdot  \bm{L}}}{p^2 - M_j^2(p^2)}~.
\end{equation}
First, it is convenient to partition the 3-momentum as
\begin{equation}
	\label{eqn:3PP}
	\bm{p} = p_L \bm{L}/L + \bm{p}_\perp~, ~~\bm{p}_\perp\cdot\bm{L} = 0~,
\end{equation}
so that $\bm{p}\cdot\bm{L} = p_LL$ and $d^3\bm{p} = d^2\bm{p}_\perp dp_L$. It is clear that the integrand of Eq. (\ref{eqn:A1PIES}) has a $p_L$ pole determined by Eq. (\ref{eqn:GEP}), which becomes in terms of $p_L$
\begin{equation}
	p_L^2 = E^2 - p_\perp^2 -m_j^2 + im_j\Gamma_j~.
\end{equation}
Since $L>0$, one can close the $p_L$ integration contour on the upper-half complex plane, and then only the $p_L$ pole in the upper-half plane contributes to the integral. Let this (positively oriented) integration contour be denoted by $C$.  (Note that a $M^2_j(p^2)$ branch cut on the $p_L$ real axis doesn't affect the integral, since it can be rotated off the axis by an appropriate choice of the principal branch.)  With the notation of Eq. (\ref{eqn:3PP}) the spatial two-point function becomes
\begin{equation}
	\label{eqn:TPF}
	\Delta_{j}(E,\bm{L}) = \int\!\! \frac{d^2 p_\perp}{(2\pi)^2}\oint_C\!\!\frac{dp_L}{2\pi} \frac{i e^{ip_LL}}{E^2 - p_\perp^2 -p_L^2 -m_j^2 +im_j\Gamma_j}~.
\end{equation}
Note that $\Delta_{j}(E,\bm{L})$ is independent of the orientation of $\bm{L}$, so $\Delta_{j}(E,\bm{L}) = \Delta_{j}(E,L)$. The physical consequence of the $p_L$ contour integration is to force the 4-momentum of the integrand (\ref{eqn:1PIES}) to be on the `pole shell' in the complex sense defined by Eq. (\ref{eqn:GEP}). We can interpret the remaining $d^2\bm{p}_\perp$ integral to be a sum over on-pole-shell transverse 3-momenta. 

Let us now perform the $p_L$ integral. In more compact notation, the propagator has a $p_L$ pole satisfying
\begin{align} 
	p_L^2(p_\perp) & =  R_j(p_\perp) + iA_j~,\label{eqn:PRPH}\\
\intertext{with}
	R_j(p_\perp) & \equiv E^2 - p_\perp^2 -m_j^2~,\notag\\ 
	A_j & \equiv m_j\Gamma_j~.
\end{align}
We do not make any assumption regarding the sign of $R_j(p_\perp)$. However, for $\Gamma_j >0$ it is clear that $\mbox{Arg}[p_L^2] \in (0,\pi]$. Therefore, defining $z\equiv A_j/R_j(p_\perp)$, it must be that
\begin{equation}
	\label{eqn:APLS}
	\mbox{Arg}[p_L^2] = \ArcTan(z) \equiv \begin{cases} \tan^{-1}|z|, ~ z > 0 \\ \pi - \tan^{-1}{|z|},~z \le 0 \end{cases} ~,
\end{equation}
where $\tan^{-1}|\cdot|: [0,\infty) \to [0,\pi/2]$. This permits us to compactly write $p_L^2(p_\perp)$ in complex polar notation. Taking a square root is now trivial, and the $p_L$ pole in the upper-half complex plane is 
\begin{equation}
	\label{eqn:EEPRP}
	p_L(p_\perp) = \Big[R^2_j(p_\perp) + A_j^2\Big]^{1/4}\exp\bigg[\frac{i}{2}\ArcTan\bigg(\frac{A_j}{R_j(p_\perp)}\bigg)\bigg]~.
\end{equation}
Applying the residue theorem to Eq. (\ref{eqn:TPF}), and observing that $p_L(p_\perp)$ is only a function of the magnitude of $\bm{p}_\perp$, we now have
\begin{equation}
	\Delta_j(E,L) = -\frac{1}{4\pi}\int_0^\infty\!\!\frac{p_\perp dp_\perp}{p_L(p_\perp)}e^{ip_L(p_\perp)L}~.
\end{equation}
Observe, furthermore, that Eq. (\ref{eqn:PRPH}) implies $p_\perp/p_L(p_\perp) = -p_L^\prime(p_\perp)$. Hence
\begin{align}
	\Delta_j(E,L) 
	& = \frac{1}{4\pi}\int_0^\infty dp_\perp p_L^\prime(p_\perp)e^{ip_L(p_\perp) L}\notag\\
	& = \frac{i}{4\pi L}e^{ip_L(0)L}~,\notag
\end{align}
as $p_L(p_\perp) \to i\infty$ when $p_\perp \to \infty$. Writing $R_j \equiv R_j(0) =  E^2 - m_j^2$, we have finally
\begin{align}
\Delta_j(E,L)  & = \frac{i}{4\pi L}\exp\bigg\{i\Big[R^2_j + A_j^2\Big]^{1/4}\exp\bigg[\frac{i}{2}\ArcTan\bigg(\frac{A_j}{R_j}\bigg)\bigg] L \bigg\}\notag\\
	& = \frac{i}{4\pi L}\exp\bigg\{\frac{i}{\sqrt{2}}\Big[\sqrt{R_j^2 + A_j^2} + R_j\Big]^{1/2}L - \frac{1}{\sqrt{2}}\Big[\sqrt{R_j^2 + A_j^2} - R_j\Big]^{1/2}L\bigg\}~.\label{eqn:AEPR}
\end{align}
In the last line we have used several trigonometric identities along with the definition of $\ArcTan$ in Eq. (\ref{eqn:APLS}).


%

\end{document}